\pdfoutput=1
\documentclass[sn-basic]{sn-jnl}

\usepackage{subfigure}
\usepackage{multirow}
\usepackage[sort&compress,numbers]{natbib}

\jyear{2021}%
\begin{document}

\title{Optimal circulant graphs as low-latency network topologies}

\author[1]{\fnm{Xiaolong} \sur{Huang}}\email{xiaolong.huang@stonybrook.edu}
\author[2]{\fnm{Alexandre} \sur{F. Ramos}}\email{alex.ramos@usp.br}
\author*[1,3]{\fnm{Yuefan} \sur{Deng}}\email{yuefan.deng@stonybrook.edu}

\affil*[1]{\orgdiv{Department of Applied Mathematics and Statistics}, \orgname{Stony Brook University}, \orgaddress{\city{Stony Brook}, \postcode{11794}, \state{New York}, \country{USA}}}

\affil[2]{\orgdiv{Escola de Artes, Ciencias e Humanidades, Nucleo de Estudos Interdisciplinares em Sistemas Complexos, Center for Translational Research in Oncology (LIM-24), Instituto do Cancer do Estado de Sao Paulo, Faculdade de Medicina}, \orgname{Universidade de Sao Paulo}, \orgaddress{\street{Avenida Arlindo Bettio 1000}, \city{Sao Paulo}, \postcode{CEP 03828-000}, \state{Sao Paulo}, \country{Brazil}}}

\affil*[3]{\orgdiv{Mathematics, Division of Science}, \orgname{New York University Abu Dhabi}, \orgaddress{\street{Saadiyat Island}, \city{Abu Dhabi}, \country{United Arab Emirates}}}

\abstract{
Communication latency has become one of the determining factors for the performance of parallel clusters. To design low-latency network topologies for high-performance computing clusters, we optimize the diameters, mean path lengths, and bisection widths of circulant topologies. We obtain a series of optimal circulant topologies of size $2^5$ through $2^{10}$ and compare them with torus and hypercube of the same sizes and degrees. We further benchmark on a broad variety of applications including effective bandwidth, FFTE, Graph 500 and NAS parallel benchmarks to compare the optimal circulant topologies and Cartesian products of optimal circulant topologies and fully connected topologies with corresponding torus and hypercube. Simulation results demonstrate superior potentials of the optimal circulant topologies for communication-intensive applications. We also find the strengths of the Cartesian products in exploiting global communication with data traffic patterns of specific applications or internal algorithms.}

\keywords{Network topology, Circulant graph, Latency, Optimization, Benchmarks}

\maketitle

\section{Introduction}\label{sec_intro}
The development of supercomputers has been advancing rapidly for the past five decades with the most recent supercomputer Fugaku culminating the Top 500 list in November 2021 at a peak speed of 537 PFlops and 7,630,848 cores \cite{top500}. Along with the growth of number of faster processors – as suggested by the Moore’s law \cite{Moore1965}, the interconnection network that can efficiently hold and connect such massive systems has been constantly on demand. Due to the power wall from the Moore’s law \cite{Moore1965}, however, the increase of processor's clock speed has hit its barrier. Consequently, the development of ever-expanding supercomputers with ever-increasing speed leans on the design of optimally interconnected networks topologies. Generally, ideal interconnection networks should satisfy small network diameter, large bisection width, topological simplicity, symmetry, engineering feasibility, and modular, expandable design, to provide maximal connectivity, scalable performance, minimal engineering complexity and least monetary cost \cite{Deng2012}. Meshes \cite{Dally2004}, tori \cite{Brightwell2006, ibm2008a, Alverson2010, Chen2011, Ajima2009, Ajima2018}, hypercubes \cite{Hayes1989}, fat trees \cite{Leiserson1985,Fu2016} and Ethernet or InfiniBand switched fabrics \cite{infiniband} have been the mainstream networks for years. Meshes have simple layout but the large diameter slows down node-to-node communications. Fat trees can realize maximum bisection width, but the diameter grows with the number of levels of switches. The torus and its variation $k$-ary $n$-cube \cite{Dally1990} have relatively smaller diameter and average distance. 3D torus is applied in Cray SeaStar \cite{Brightwell2006} and IBM Blue Gene/P \cite{ibm2008a}, while modified 3D torus formulates Cray Gemini interconnect \cite{Alverson2010}. 5D torus is incorporated in IBM Blue Gene/Q \cite{Chen2011}. Hybrid 6D mesh/torus Tofu interconnect is integrated in K and post-K supercomputers \cite{Ajima2009,Ajima2018}. Recently, Dragonfly topology \cite{Kim2008} among hierarchical topologies with small diameter has been deployed into Aries and Slingshot interconnects \cite{Faanes2012,DeSensi2020}.

In this manuscript, we propose the optimal circulant topologies with minimal diameter and mean path length (MPL) and maximum bisection width as promising low-latency network topologies by a home-grown and highly efficient parallel exhaustive search algorithm to generate and optimize circulant topologies. We obtain a set of optimal circulant topologies of size $2^5$ through $2^{10}$. By comparing with corresponding torus and hypercube of the same sizes and degrees, we show remarkable improvement in graph properties by the optimal circulant topologies and the Cartesian products of optimal circulant topologies and fully connected topologies. The benchmarking results by simulation also demonstrate significant performance enhancement for communication-intensive applications. In the meantime, the Cartesian products of optimal circulant topologies and fully connected topologies show potential in balancing the influence of global communication with data traffic patterns of specific applications or internal algorithms. We discuss related work in Sect. \ref{sec_rel_w} and present the parallel optimization algorithm for circulant topologies in Sect. \ref{sec_discover}. Section \ref{sec_compare} examines graph properties of optimal circulant topologies and Cartesian products by comparative analysis with torus and hypercube. Section \ref{sec_sim} shows simulated benchmark performance and detailed analysis. We conclude the manuscript in Sect. \ref{sec_conclude}.

\section{Related work}\label{sec_rel_w}
The optimization of interconnection networks is essentially to reduce the node-to-node communication latency, which is directly indicated by the diameter and mean path length (MPL) of the topology. The idea of optimizing MPL can be dated back to Cerf et al., who first calculated the lower bound of MPL given arbitrary graph size and degree \cite{Cerf1974}. Zhang et al. modified torus by adding bypass links to construct interlaced bypass torus (iBT) \cite{Zhang2011,Zhang2012_ibt}, designed efficient broadcast algorithms on iBT \cite{Zhang2012_ibt_bcast} and evaluated its performance by building a re-configurable cluster and system \cite{Zhang2015}. Feng et al. further explored 6D mesh/iBT on custom routing algorithms \cite{Feng2013} and performance evaluation by simulation \cite{Feng2012}. Deng et al. implemented parallel exhaustive search of regular graphs \cite{Xu2019} and random search with rotational symmetry to obtain small and large (near) optimal topologies and evaluated the performance by benchmarking both on real clusters and simulation platform \cite{Deng2020}. More small optimal topologies with symmetries and other special properties were produced and presented in a structured table \cite{Zhang2019}. Distributed shortcut networks targeting the diameter and cable length trade-off \cite{Truong2017} and host-switch graphs designed by minimizing diameter and/or MPL using randomized heuristics \cite{Yasudo2019} were also introduced and benchmarked by simulation.

In addition, network topologies with additional symmetry properties provide advantages in routing and load balancing \cite{Dally2004} and also in design and engineering simplicity. Recent research on constructing supercomputer network utilizing Lie algebra symmetry \cite{Deng2012,Sabino2018} and optimizing MPL by random or heuristic algorithms with rotational symmetry constraints \cite{Deng2020,Nakao2021} have demonstrated the importance of the symmetry perspective in network design. The total number of symmetries of a graph calculated as its automorphism group size has also been included in further optimization of smaller topologies \cite{Zhang2019}. Vertex-symmetric topologies are one class of highly symmetric topologies, where any two vertices are equivalent under self mappings that preserve adjacency \cite{Dally2004}. Torus consisting of equal-sized rings, hypercube and dragonfly are notable examples of vertex-symmetric topologies that have been widely applied in current mainstream high-performance computing networks as introduced in Sect. \ref{sec_intro}.

Among vertex-symmetric topologies, circulant topologies, also known as multi-loop networks, have been extensively explored in both theory and applications \cite{Bermond1995, Hwang2003, MONAKHOVA2012}. The lower bounds of diameter and MPL for circulant topologies have been estimated in theory, which can be achieved especially for degree-4 (double-loop) circulants \cite{Hwang2003}. However, for arbitrary degree, the existence of such optimal circulant topologies that reach theoretical lower bounds of diameter and MPL still remains open.

The synthesis of arbitrary-degree circulant topologies with the smallest diameter and/or MPL has been conducted mostly through heuristic algorithms \cite{MONAKHOVA2012}. A related problem is that given fixed diameter and degree, finding large circulant topologies by either theoretical construction or computer search, which has also been investigated \cite{FeriaPurn2014, Bevan2017}. However, to the best of our knowledge, there are only a few parallel algorithms for the synthesis and optimization of arbitrary-degree circulant topologies, and even less work on the performance evaluation of circulant topologies as communication networks. Our optimization method of efficient parallel exhaustive search provides another approach in finding the definitive optimal circulant topology. We further select the optimal circulant topology with the largest bisection width to add to its potential in performance enhancement. Moreover, our simulation results bring insight into the actual performance of the optimal circulant topologies for computational applications.

\section{Discovery of optimal circulant topologies}\label{sec_discover}

\begin{figure}[htbp!]
	\centering
	\subfigure[$(16,4)$ Optimal Circulant, Jumps = 1,6]{
		\includegraphics[width=0.45\textwidth]{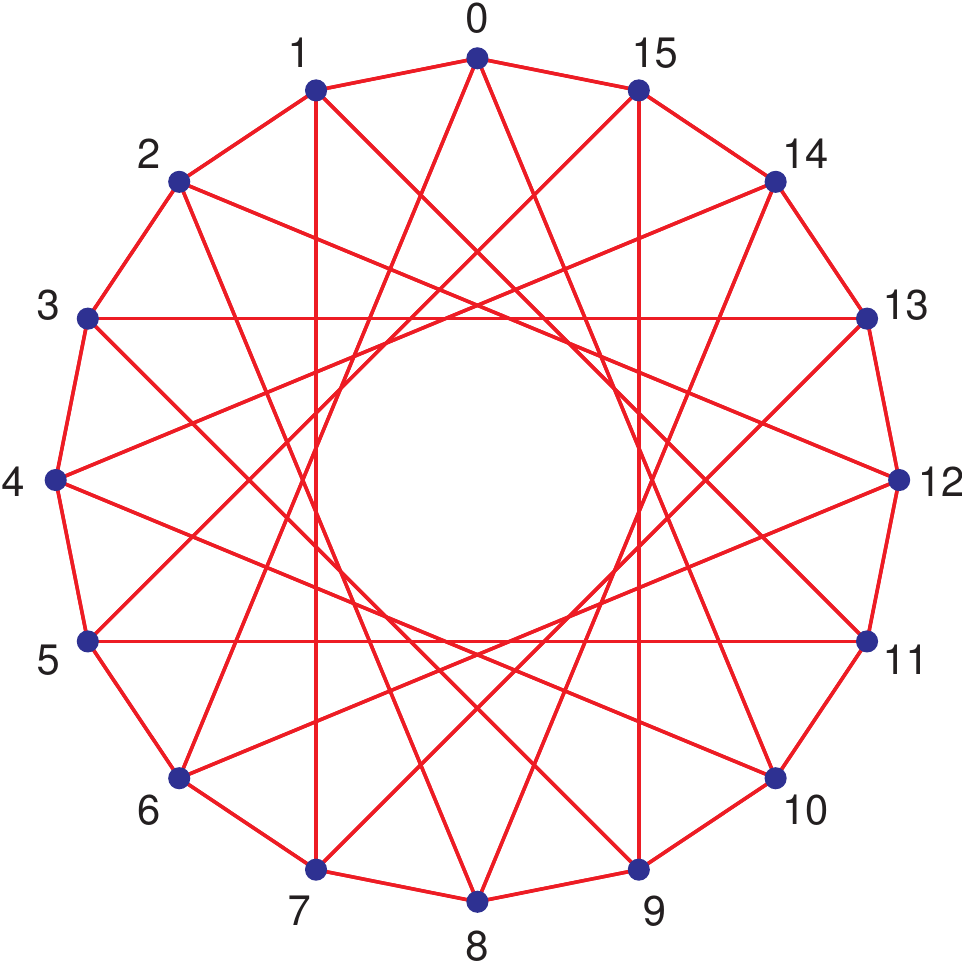}}
	\qquad
	\subfigure[$(16,5)$ Optimal Circulant, Jumps = 1,3,8]{
		\includegraphics[width=0.45\textwidth]{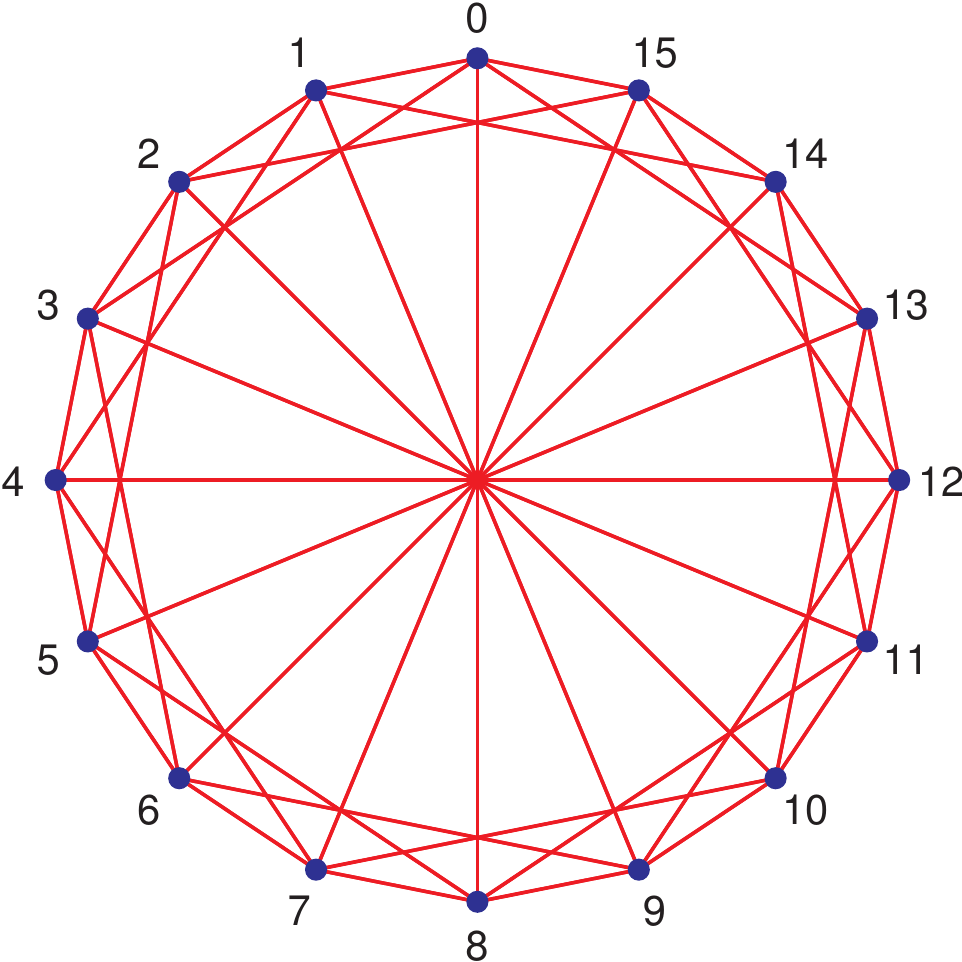}}
	\caption{16-vertex Optimal circulant graphs}
	\label{16_opt_circ}
\end{figure}

An $N$-vertex circulant graph \cite{gross2013handbook} is defined by a jump set $S=\{s \vert 1 \leq s \leq \lfloor{N/2}\rfloor\}$, where each vertex $i=0, 1, ..., N-1$ is connected to vertices $i\pm s \pmod{N}$ for every jump $s$. A degree-$k$ circulant graph has $\lceil{k/2}\rceil$ number of jumps. In particular, the rings and complete graphs are both circulant, when $S=\{1\}$ and $S=\{1, 2, ..., \lfloor{N/2}\rfloor\}$ respectively. Examples of 16-vertex circulant graphs are shown in Fig. \ref{16_opt_circ}, which are optimal as we discovered.

We search optimal circulant graphs of minimal diameter (D), minimal mean path length (MPL), and maximal bisection width (BW). To enumerate all $N$-vertex degree-$k$ circulant graphs, denoted by $(N,k)$, is equivalently to compute all combinations of choosing $\lfloor{k/2}\rfloor$ jumps from 1 to $\lfloor{(N-1)/2}\rfloor$. Jump $N/2$ is excluded as the jump set $S$ contains $N/2$ if and only if $k$ is odd. Since we require the circulant graphs to be connected, by theorem of Boesch and Tindell \cite{Boesch1984}, a circulant graph is connected if and only if $gcd(N,S)=1$. Thus, when $N$ is a power of 2, there must be an odd jump. Such an odd jump is also relatively prime to $N$, which can be mapped to jump 1 in an equivalent jump set by Adam isomorphism of the circulant graph \cite{Boesch1984}. As a result, for $N$ equal power of 2, we can set one of the jumps to 1 and reduce the exhaustive search to choosing $\lfloor{k/2}\rfloor-1$ jumps from 2 to $\lfloor{(N-1)/2}\rfloor$. For instance, the numbers of $(1024,10)$ and $(1024,15)$ circulant graphs to optimize can be considerably reduced from $\sim 3 \cdot 10^{11}$ to $\sim 3 \cdot 10^{9}$ and from $\sim 2 \cdot 10^{15}$ to $\sim 2 \cdot 10^{13}$ respectively.

To effectively perform the exhaustive search of optimal circulant graphs, we adapt the co-lexicographic ordering algorithm of combinations in the FXT library, which is a highly efficiently library in combinatorial generation \cite{arndt2010matters, fxt}. One key feature of the algorithm is that the generation of each combination only depends on the previous combination in co-lexicographic order. Therefore, the generation process can be reduced to a minimum number of operations which makes it as fast as possible. Moreover, such feature enables the parallelization of the total combinatorial generation by incorporating with an unranking algorithm \cite{ruskey2003combinatorial}. In the co-lexicographic ordering, each combination corresponds to a unique ordinal number which is called the rank of the combination. The unranking algorithm maps a rank back to its corresponding combination in co-lexicographic order, which allows the generation to start and end at arbitrary ranks.

Utilizing the above combinatorial techniques, given $(N,k)$ as input, we design the parallel exhaustive search algorithm to start by counting the total number of $(N,k)$ circulant graphs to optimize, and divide the number evenly among all processors. Then each processor unranks its start rank, generates circulant graphs by enumerating through combinations in co-lexicographic order while filtering by finding the ones with minimal diameter and then minimal MPL, and stops at its end rank. For the calculation of diameter and MPL, we run breadth-first search only once on each graph since circulant graphs are vertex-symmetric. For $N$ equal power of 2, we use bit arrays and bit shifts to mark traversed vertices in the implementation of breadth-first search to further improve its speed. Finally, the root processor finds the minimal diameter and MPL among all processors and filters once more to obtain the optimal circulant graphs. Hence our parallel exhaustive search algorithm has perfectly balanced workloads and highly efficient implementations of both combinatorial generation and calculation of diameter and MPL. We execute the parallel exhaustive search on the SeaWulf supercomputer at Stony Brook University. The processing speed on compute nodes with Intel Xeon Gold 6148 CPUs can reach $\sim 10^{5}$ graphs per second per core, which makes it practical to optimize large high-degree circulant graphs even up to $(1024,15)$.

When there are multiple optimal circulant graphs with minimal diameter and MPL, we filter further by optimizing bisection width. To compute the bisection width of a graph, we use the KaHIP program which can achieve strictly balanced bisection as required \cite{Sanders2013}. We present the discovered optimal circulant graphs (in Table \ref{table_opt_circ}) which have minimal diameters and MPLs and maximum bisection widths filtered in such specific order. We believe they serve as promising candidates for interconnection network topologies.

\begin{table}[htbp!]
\begin{center}
\caption{Optimal circulant graphs discovered by parallel exhaustive search}
\label{table_opt_circ}
\begin{tabular}{rrl}
	\toprule
	$N$ & $k$ & Jumps \\
	\midrule
	\multirow{2}{*}{32} &4 &1,7\\
	&5 &1,6,16\\
	\midrule
	\multirow{3}{*}{64} &4 &1,14\\
	&5 &1,7,32\\
	&6 &1,4,25\\
	\midrule
	\multirow{2}{*}{128} &6 &1,8,54\\
	&7 &1,12,30,64\\
	\midrule
	\multirow{3}{*}{256} &6 &1,47,122\\
	&7 &1,13,33,128\\
	&8 &1,9,74,103\\
	\midrule
	\multirow{2}{*}{512} &8 &1,15,56,149\\
	&9 &1,23,31,119,256\\
	\midrule
	\multirow{2}{*}{1024} &8 &1,144,258,276\\
	&10 &1,38,70,393,481\\
	\bottomrule
\end{tabular}
\end{center}
\end{table}

\section{Graph analysis of optimal circulant topologies}\label{sec_compare}
We analyze the graph properties of optimal circulant topologies by comparing with torus and hypercube as well as Cartesian products of one smaller optimal circulant topology and the 4-vertex fully connected topology. We first calculate and present the diameter, MPL and bisection width of the three categories of topologies in Table \ref{graph_prop_tab}. For each $N$ from $2^5$ to $2^{10}$, single optimal circulant topologies are denoted as Optimal Circulant, while Cartesian products of optimal circulant topologies are denoted as Optimal Circulant Product. For tori and Cartesian products, the product components are placed from small to large starting on the right.

\begin{table}[htbp!]
\begin{center}
\caption{Graph properties of optimal circulant (product) topologies and tori/hypercubes}
\label{graph_prop_tab}
\begin{tabular}{rrlrcr}
	\toprule
	$N$ & $k$ & Topology & D & MPL & BW \\
	\midrule
	\multirow{5}{*}{32} &\multirow{2}{*}{4} &Torus 2D (8,2)$\times$(4,2) &6 &3.10 &8\\
	&&Optimal Circulant &4 &2.71 &16\\
	\cline{2-6}
	&\multirow{3}{*}{5} &Optimal Circulant &3 &2.29 &20\\
	&&Optimal Circulant Product (8,2)$\times$(4,3) &5 &2.84 &8\\
	&&Hypercube 5D &5 &2.58 &16\\
	\midrule
	\multirow{5}{*}{64} &\multirow{2}{*}{4} &Torus 2D (8,2)$\times$(8,2) &8 &4.06 &16\\
	&&Optimal Circulant &6 &3.78 &22\\
	\cline{2-6}
	&\multirow{3}{*}{6} &Optimal Circulant &4 &2.60 &48\\
	&&Optimal Circulant Product (16,3)$\times$(4,3) &5 &3.24 &16\\
	&&Hypercube 6D &6 &3.05 &32\\
	\midrule
	\multirow{5}{*}{128} &\multirow{2}{*}{6} &Torus 3D (8,2)$\times$(4,2)$\times$(4,2) &8 &4.03 &32\\
	&&Optimal Circulant &5 &3.34 &76\\
	\cline{2-6}
	&\multirow{3}{*}{7} &Optimal Circulant &4 &3.02 &100\\
	&&Optimal Circulant Product (32,4)$\times$(4,3) &5 &3.40 &64\\
	&&Hypercube 7D &7 &3.53 &64\\
	\midrule
	\multirow{5}{*}{256} &\multirow{2}{*}{6} &Torus 3D (8,2)$\times$(8,2)$\times$(4,2) &10 &5.02 &64\\
	&&Optimal Circulant &6 &4.25 &120\\
	\cline{2-6}
	&\multirow{3}{*}{8} &Optimal Circulant &5 &3.33 &234\\
	&&Optimal Circulant Product (64,5)$\times$(4,3) &5 &3.78 &128\\
	&&Hypercube 8D &8 &4.02 &128\\
	\midrule
	\multirow{5}{*}{512} &\multirow{2}{*}{8} &Torus 4D (8,2)$\times$(4,2)$\times$(4,2)$\times$(4,2) &10 &5.01 &128\\
	&&Optimal Circulant &6 &4.04 &392\\
	\cline{2-6}
	&\multirow{3}{*}{9} &Optimal Circulant &5 &3.73 &472\\
	&&Optimal Circulant Product (128,6)$\times$(4,3) &6 &4.07 &304\\
	&&Hypercube 9D &9 &4.51 &256\\
	\midrule
	\multirow{5}{*}{1024} &\multirow{2}{*}{8} &Torus 4D (8,2)$\times$(8,2)$\times$(4,2)$\times$(4,2) &12 &6.01 &256\\
	&&Optimal Circulant &7 &4.88 &624\\
	\cline{2-6}
	&\multirow{3}{*}{10} &Optimal Circulant &5 &4.07 &1036\\
	&&Optimal Circulant Product (256,7)$\times$(4,3) &6 &4.52 &624\\
	&&Hypercube 10D &10 &5.00 &512\\
	\bottomrule
\end{tabular}
\end{center}
\end{table}

\begin{figure}[hbtp!]
	\centering
	\includegraphics[width=1\textwidth]{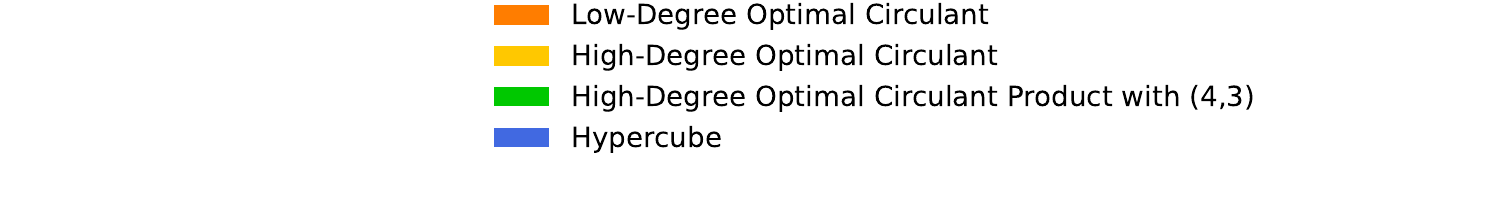}
	\caption{Color scheme to represent different topology categories in all bar plots}
	\label{fig_legend}
\end{figure}

\begin{figure}[htbp!]
	\centering
	\subfigure[Graph properties average (inverse) ratios]{
		\includegraphics[width=0.7\textwidth]{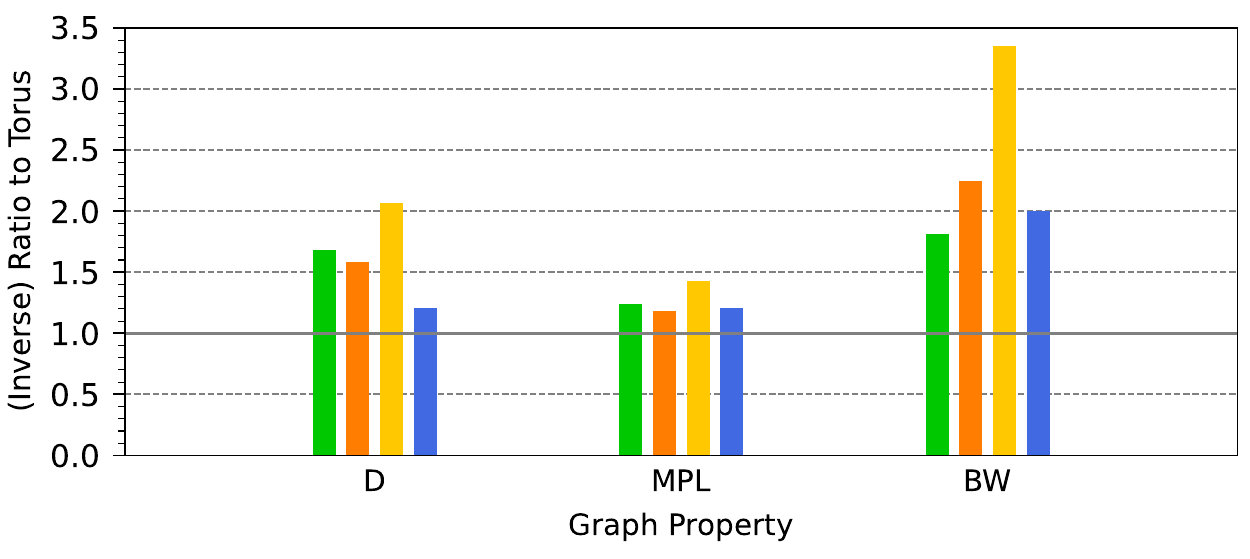}}
	\\\vspace*{0.3\baselineskip}
	\subfigure[D inverse ratios]{
		\includegraphics[width=0.7\textwidth]{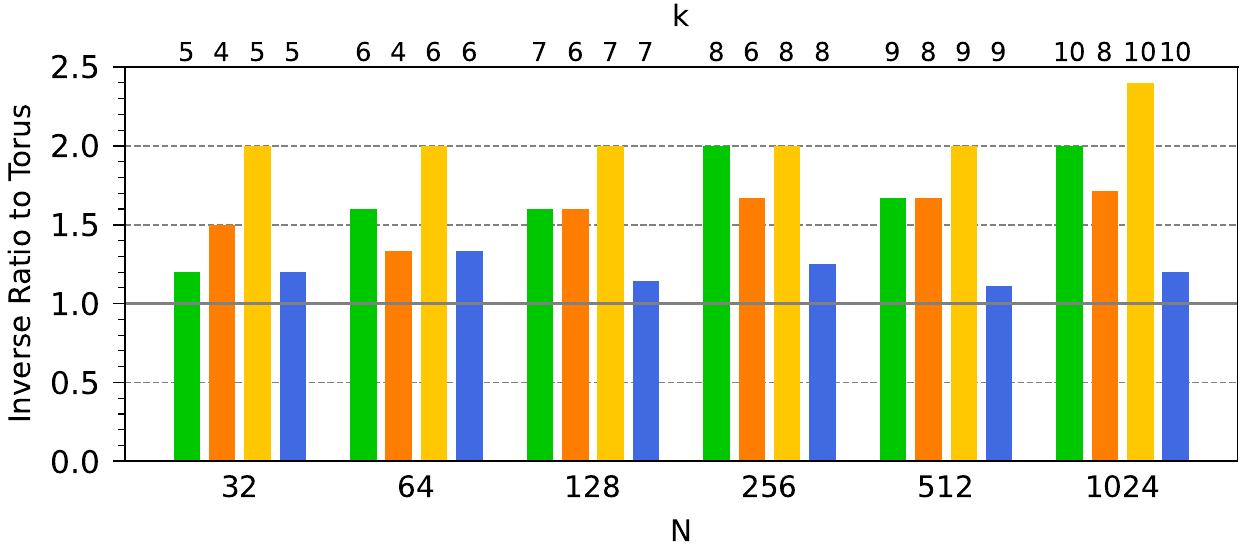}}
	\\\vspace*{0.3\baselineskip}
	\subfigure[MPL inverse ratios]{
		\includegraphics[width=0.7\textwidth]{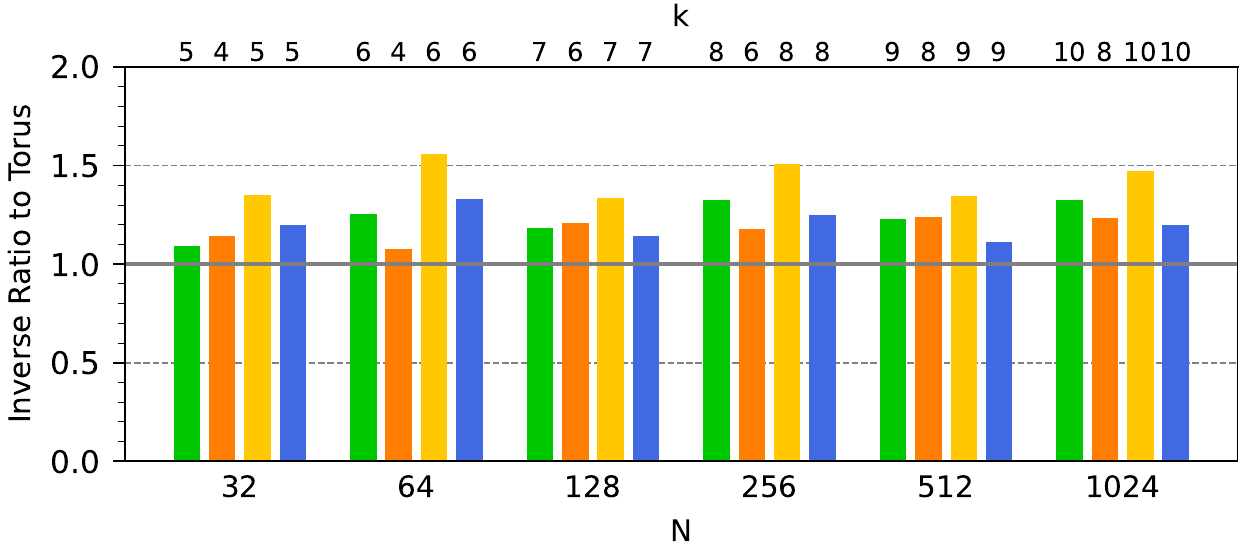}}
	\\\vspace*{0.3\baselineskip}
	\subfigure[BW ratios]{
		\includegraphics[width=0.7\textwidth]{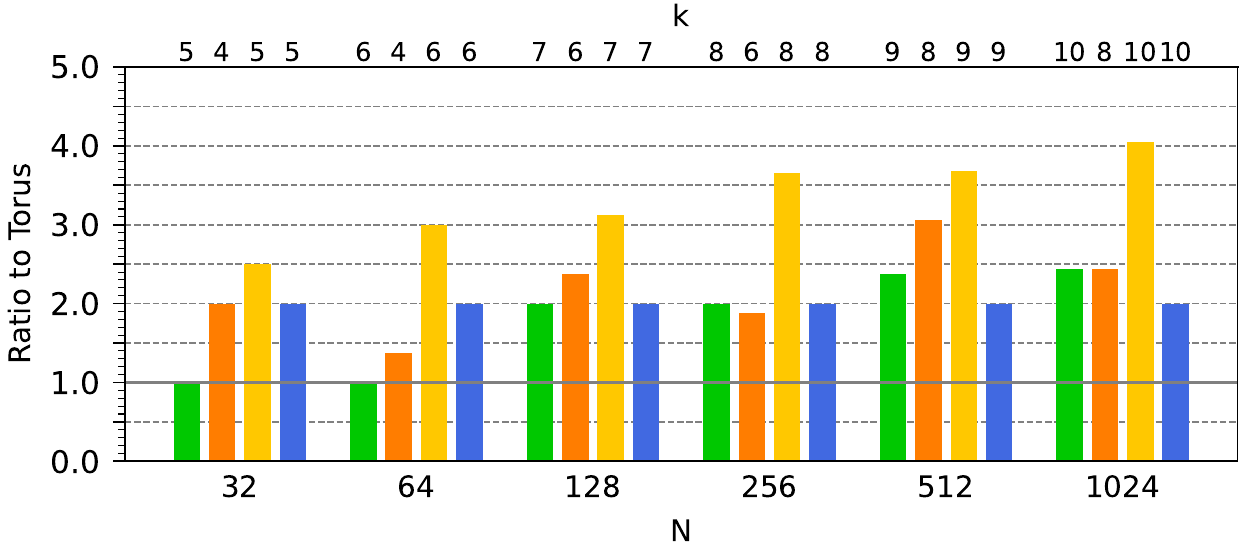}}
	\caption{Graph properties measured in (inverse) ratios to torus}
	\label{graph_prop_fig}
\end{figure}

Moreover, for a given graph size we further normalize the graph properties in Table \ref{graph_prop_tab} by comparing all topologies with torus and compute the graph property ratios. For diameter and MPL, the ratios are inverted as $\text{D}_\text{Torus}/\text{D}_\text{Topology}$ and $\text{MPL}_\text{Torus}/\text{MPL}_\text{Topology}$. For bisection width, the ratio stays as $\text{BW}_\text{Topology}/\text{BW}_\text{Torus}$. All (inverse) ratios are shown in Fig. \ref{graph_prop_fig}b-d by bar plots, such that higher bars always represent better graph properties. In the end, for each graph property and graph category, we show the average (inverse) ratios across graph size $N$ in Fig. \ref{graph_prop_fig}a. We use low-degree and high-degree to denote the two different graph degrees for each $N$ as in Table \ref{graph_prop_tab}, and use consistent color scheme as in Fig. \ref{fig_legend} to represent different topology categories in all bar plots Figs. \ref{graph_prop_fig}-\ref{fig_npb2}.

Table \ref{graph_prop_tab} and Fig. \ref{graph_prop_fig} show that the optimal circulant topologies and Cartesian products always have smaller diameter, MPL and larger bisection width than corresponding torus. The low-degree optimal circulant topologies and Cartesian products have average diameter inverse ratios 1.58/1.68 and average MPL inverse ratios 1.18/1.24 respectively comparing with torus, indicating 37\%/40\% smaller diameter and 15\%/19\% smaller MPL in average. Meanwhile, the high-degree optimal circulant topologies have further smaller diameter and MPL, 52\% and 30\% decrease in average comparing with torus. For bisection width, the low-degree optimal circulant topologies and Cartesian products have clear advantages over torus, and almost as good as hypercube, while the high-degree optimal circulant topologies can achieve up to 235\% average increase. We also observe that for the optimal circulant topologies and Cartesian products, both the diameter inverse ratios and bisection width ratios are increasing along with the topology size $N$ with slight fluctuation, while the MPL inverse ratios mostly stay around its average.

\section{Benchmarks of optimal circulant topologies}\label{sec_sim}
\subsection{Simulation platform and network routing}
To investigate the performance of optimal circulant topologies, we simulate the topologies listed in Table \ref{graph_prop_tab} on SimGrid (version 3.26) \cite{Casanova2014}. SimGrid offers flexible and accurate parallel simulation platform, especially the SMPI interface which enables simulation of MPI applications \cite{Casanova2014}. We configure the parameters in SimGrid by setting 100 GFlops computational speed per cluster node, 10 Gbps network bandwidth and 40 ns latency per link. We utilize the built-in implementation of MVAPICH2 in SimGrid as the MPI library. All the simulations are run on the SeaWulf supercomputer at Stony Brook University.

For network routing, we use static shortest-path routing with full routing table for all simulated topologies with a custom-designed vertex-symmetric routing for the optimal circulant topologies. To determine the routing table, we first apply breadth-first search once and find the initial routes from node 0 to all the other nodes. Then we add $i$ ($\bmod\  N$) to all the nodes on the initial routes to form the routes from node $i$ to all the other nodes. In this way we can construct the full routing table for the optimal circulant topologies. For torus, hypercube, and Cartesian products, we implement the widely used dimension-order routing which achieves shortest distances between nodes in Cartesian product topologies \cite{Dally2004}.

\subsection{Benchmark results and analysis}
\begin{figure}[hbtp!]
	\centering
	\includegraphics[width=1\textwidth]{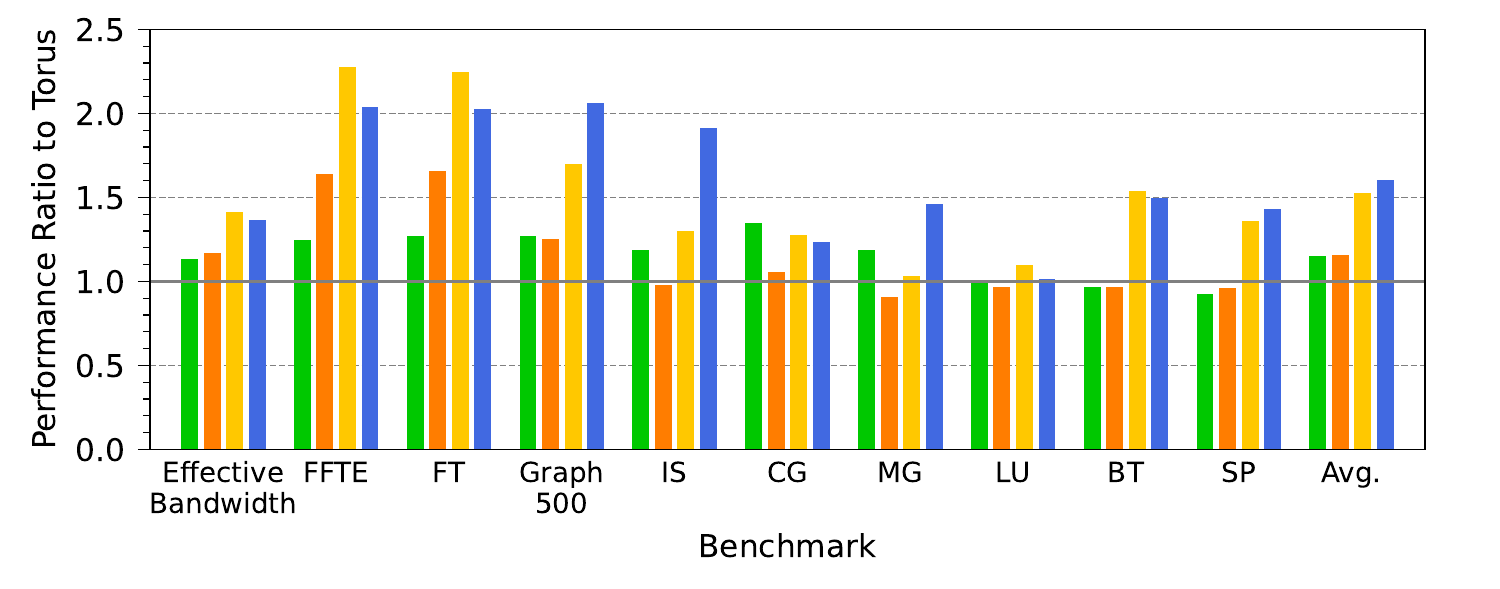}
	\caption{Average performance measured in ratios to torus}
	\label{fig_sim_avg}
\end{figure}

The following benchmark programs are simulated to examine and compare the performance of network topologies: effective bandwidth (b\_eff) \cite{effec, Konigesa}, FFTE \cite{ffte, Takahashi2000}, Graph 500 \cite{graph500, murphy2010introducing} and the NAS Parallel Benchmarks (NPB) \cite{npb, Bailey1991} consisting of FT, IS, CG, MG, LU, BT and SP. We run each benchmark on all topologies listed in Table \ref{graph_prop_tab} and evaluate the performance by ratio of processing speeds to torus of the same size. The performance ratio is calculated as $S_\text{Topology}/S_\text{Torus}$, where $S$ is the average speed over multiple runs reported by the benchmark. All performance ratios are demonstrated by bar plots in Figs. \ref{fig_sim_avg}-\ref{fig_npb2}. The terms low-degree and high-degree are used to denote the two different graph degrees for each $N$. The same color scheme as in Fig. \ref{fig_legend} is used to represent different topology categories in all bar plots. Like the comparative analysis of graph properties in Section \ref{sec_compare}, we also present in Fig. \ref{fig_sim_avg} the average performance ratios across topology size $N$ for each benchmark.

Figure \ref{fig_sim_avg} shows that comparing with torus, the low-degree and high-degree optimal circulant topologies and Cartesian products achieve average performance gains of 15\%, 52\% and 15\% respectively over all simulated benchmarks. The performance ratios are much higher especially on effective bandwidth, FFTE, NPB FT and Graph 500, while reaching higher or equal average performance for the other NPB benchmarks. In particular, the high-degree optimal circulant topologies are capable of enhancing the performance up to 128\% and 124\% in average on FFTE and NPB FT respectively, even 24\% and 21\% higher than hypercube. We present more detailed analysis on the benchmarks and simulation results in the rest of this section.

\subsubsection{Effective bandwidth}
Effective bandwidth (b\_eff, version 3.6.0.1) \cite{effec, Konigesa} measures the total network bandwidth over multiple ring and random communication patterns. It also compares three different communication methods: MPI\_Sendrecv, MPI\_Alltoallv and non-blocking MPI\_Irecv and MPI\_Isend with MPI\_Waitall, and reports the maximum bandwidth of all methods. To fairly compare the performance of optimal circulant topologies with torus, we modify the benchmark to run over random communication patterns only and collect the reported bandwidth at maximum message size of 1 MB per processor.

\begin{figure}[htbp!]
	\centering
	\includegraphics[width=0.7\textwidth]{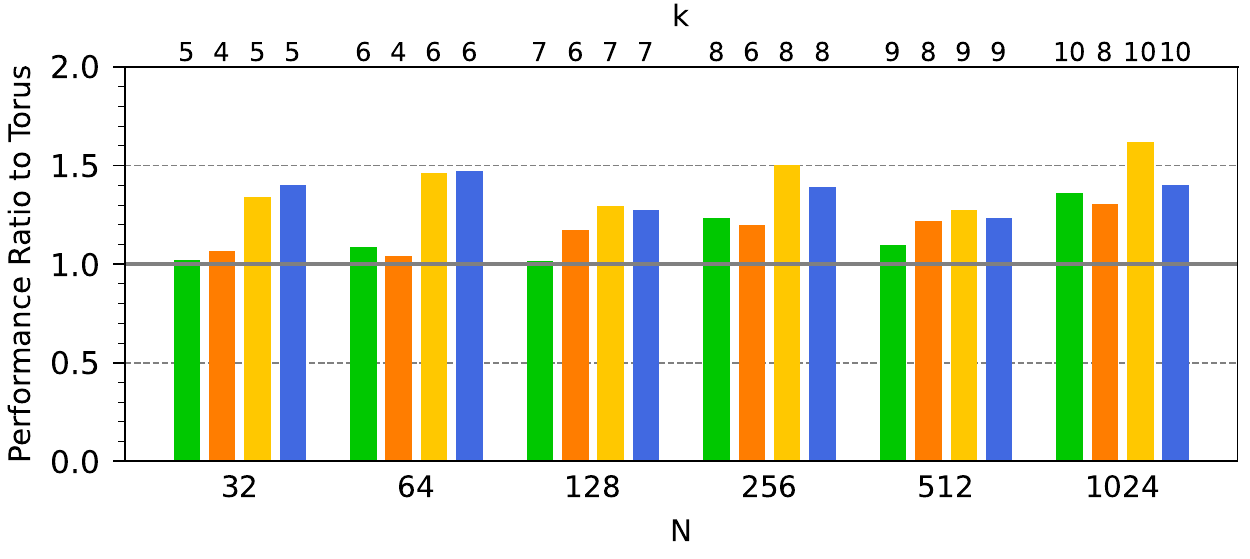}
	\caption{Performance ratios to torus on effective bandwidth}
	\label{fig_beff}
\end{figure}

The effective bandwidth performance ratios to torus are plotted in Fig. \ref{fig_beff}. The high-degree optimal circulant topologies substantiate 42\% average performance gain, reaching 62\% at $N=1024$ which is even 22\% higher that hypercube. The low-degree optimal circulant topologies and Cartesian products show higher performance gain starting with larger topology size from $N=256$, with 17\%/14\% in average and as high as 31\%/34\% respectively at $N=1024$, almost equal to hypercube. Since effective bandwidth essentially measures cross-network traffic, optimal circulant topologies with small diameter and MPL and large bisection width therefore can significantly improve the performance.

\subsubsection{FFTE}\label{sec_ffte}
FFTE (version 7.0) \cite{ffte, Takahashi2000} is simulated, which requires global all-to-all data transpositions across the network. We perform the parallel 1D FFTE routine with transform array lengths ranging from $2^{10}$ to $2^{31}$, equal to total transform array sizes from 16 KB to 32 GB. Then we collect the sustained computational speed at the maximum transform array length.

\begin{figure}[htbp!]
	\centering
	\includegraphics[width=0.7\textwidth]{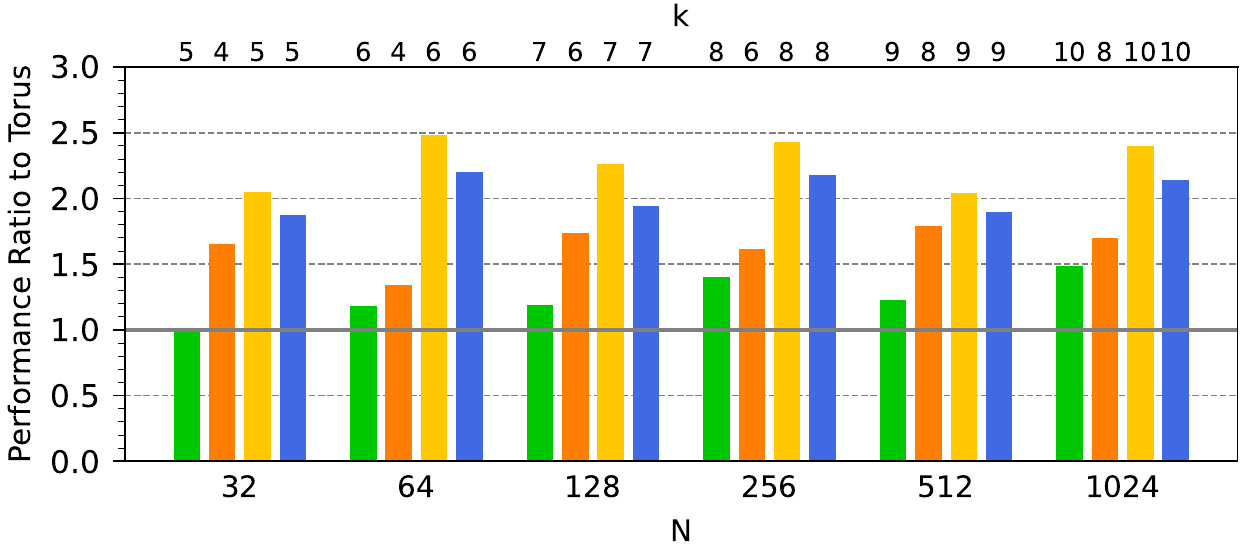}
	\caption{Performance ratios to torus on FFTE}
	\label{fig_ffte}
\end{figure}

Figure \ref{fig_ffte} shows the FFTE performance ratios to torus. The low-degree and high-degree optimal circulant topologies and Cartesian products have slightly fluctuating performance ratios, maintaining average performance gains of 64\%, 128\% and 25\% respectively. Moreover, the high-degree optimal circulant topologies outperform hypercube for every topology size, adding to 24\% higher average performance gain. The high performance gains result from the performance dependence of FFTE on global communication, which relies heavily on the optimization of circulant network topology.

\subsubsection{Graph 500}\label{sec_graph500}
The Graph 500 (version 2.1.4) \cite{graph500, murphy2010introducing} tests graph search and shortest path algorithms on an tremendously large undirected graph distributed among all processors. It evaluates large-scale data-intensive performance for supercomputers. The processing speed is reported as mean TEPS, i.e. traversed edges per second. Due to implementation issues with SimGrid, we use the previous version 2.1.4 and the replicated version of parallel breadth-first search. We choose the scale of 29 with edge factor 4, generating an initial unweighted graph of 36 GB.

\begin{figure}[htbp!]
	\centering
	\includegraphics[width=0.7\textwidth]{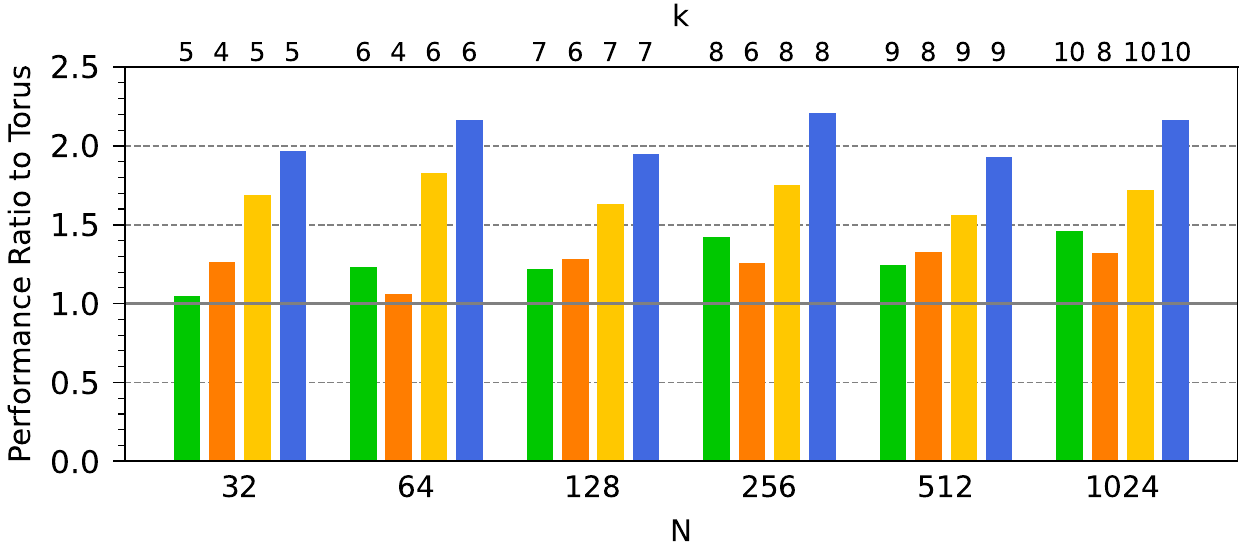}
	\caption{Performance ratios to torus on Graph 500}
	\label{fig_graph500}
\end{figure}

The performance ratios to torus for Graph 500 are shown in Fig. \ref{fig_graph500}. The low-degree and high-degree optimal circulant topologies and Cartesian products can achieve average performance gains of 25\%, 70\% and 27\% respectively. The global communication involved in Graph 500 makes the optimal circulant topologies suitable for enhancing the performance. We also note that hypercube brings top performance for Graph 500, which shows that the extremely symmetric topology structure coupled with proper routing may compensate its other disadvantages by matching with traffic patterns and internal MPI algorithms.

\subsubsection{NAS parallel benchmarks (NPB)}

\begin{figure}[htbp!]
	\centering
	\subfigure[NPB FT]{
		\includegraphics[width=0.7\textwidth]{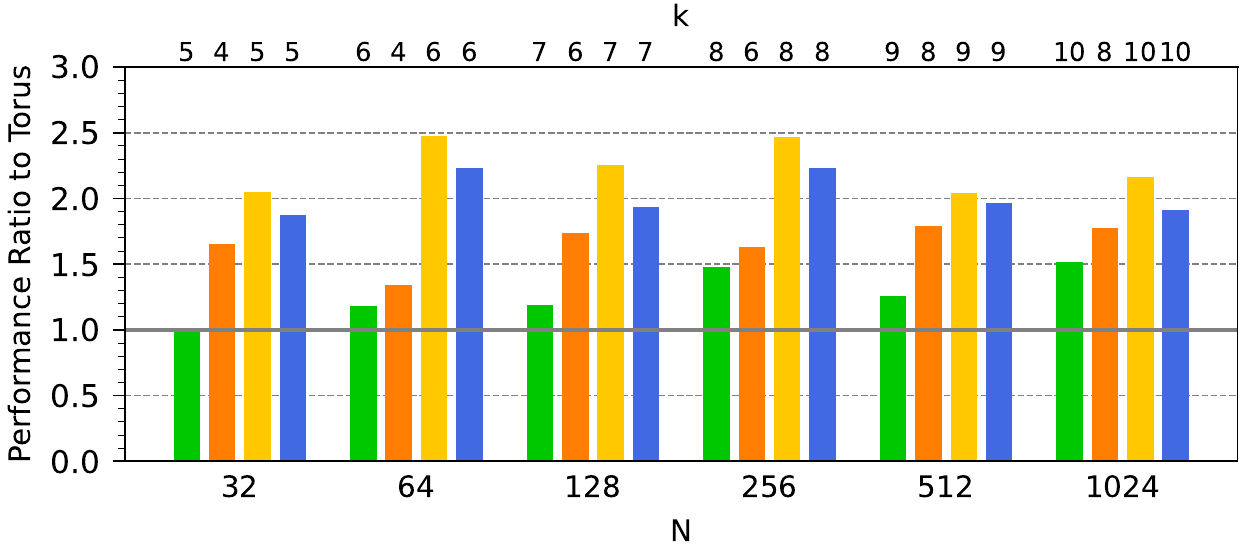}}
	\\\vspace*{0.3\baselineskip}
	\subfigure[NPB IS]{
		\includegraphics[width=0.7\textwidth]{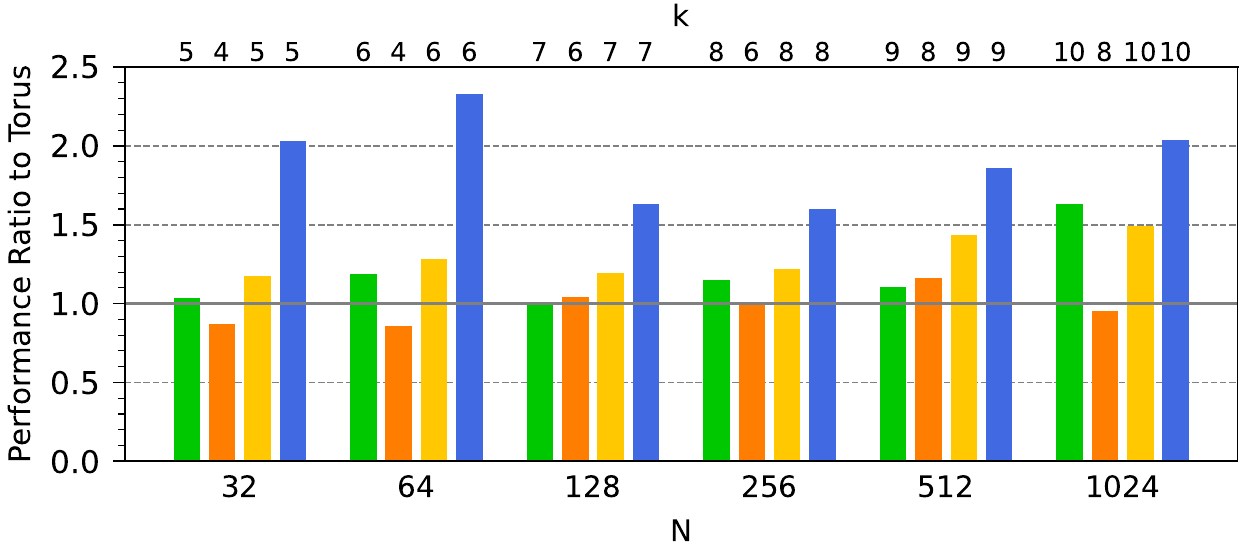}}
	\\\vspace*{0.3\baselineskip}
	\subfigure[NPB CG]{
		\includegraphics[width=0.7\textwidth]{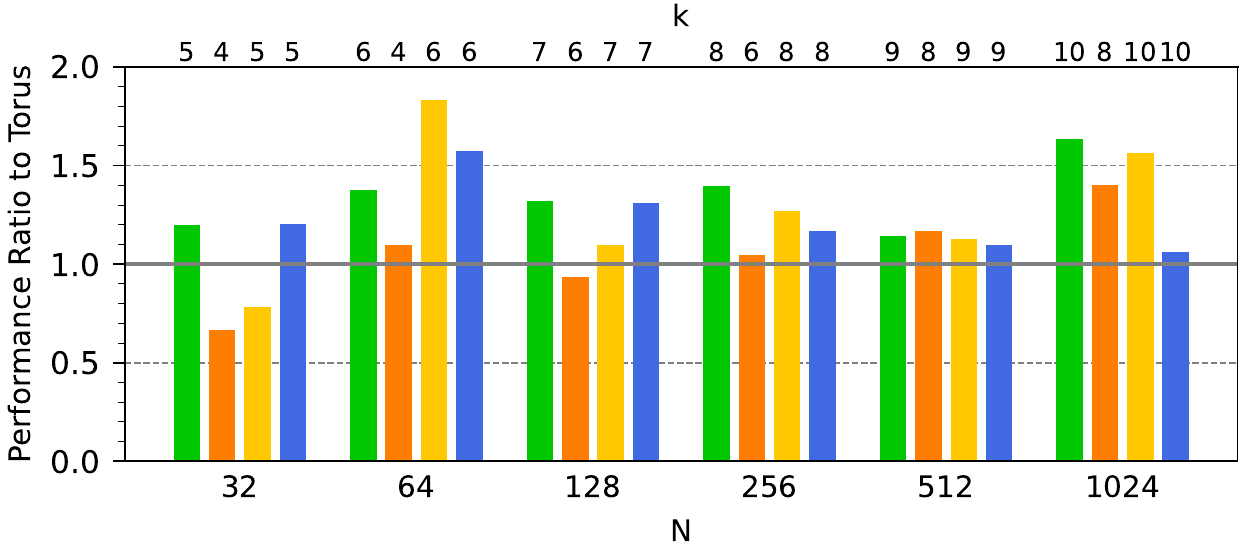}}
	\\\vspace*{0.3\baselineskip}
	\subfigure[NPB MG]{
		\includegraphics[width=0.7\textwidth]{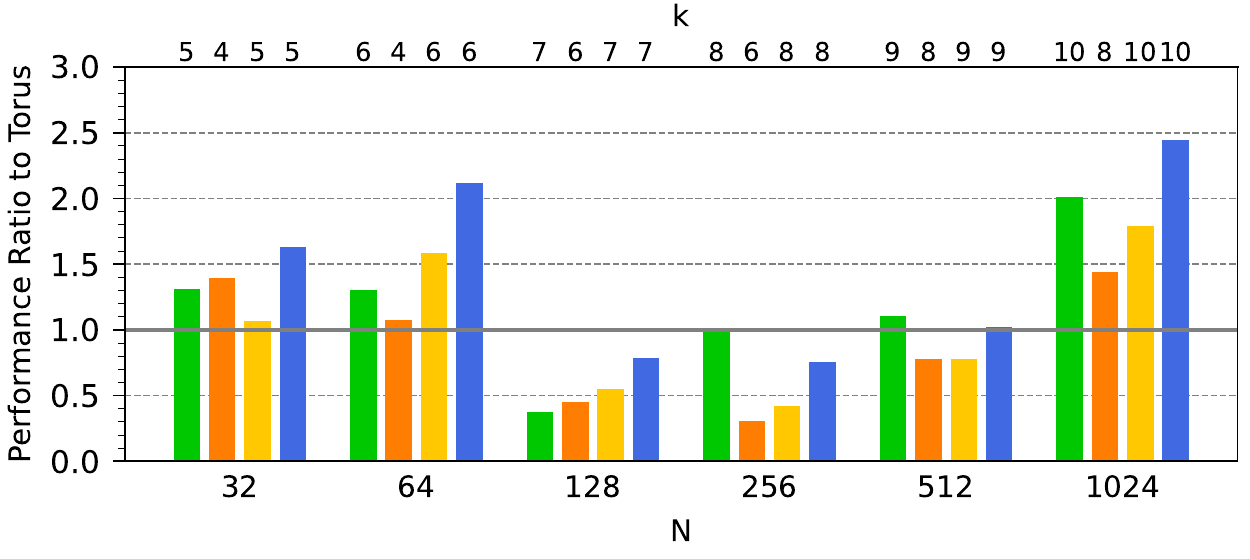}}
	\caption{Performance ratios to torus on NPB kernels}
	\label{fig_npb1}
\end{figure}

\begin{figure}[htbp!]
	\centering
	\subfigure[NPB LU]{
		\includegraphics[width=0.7\textwidth]{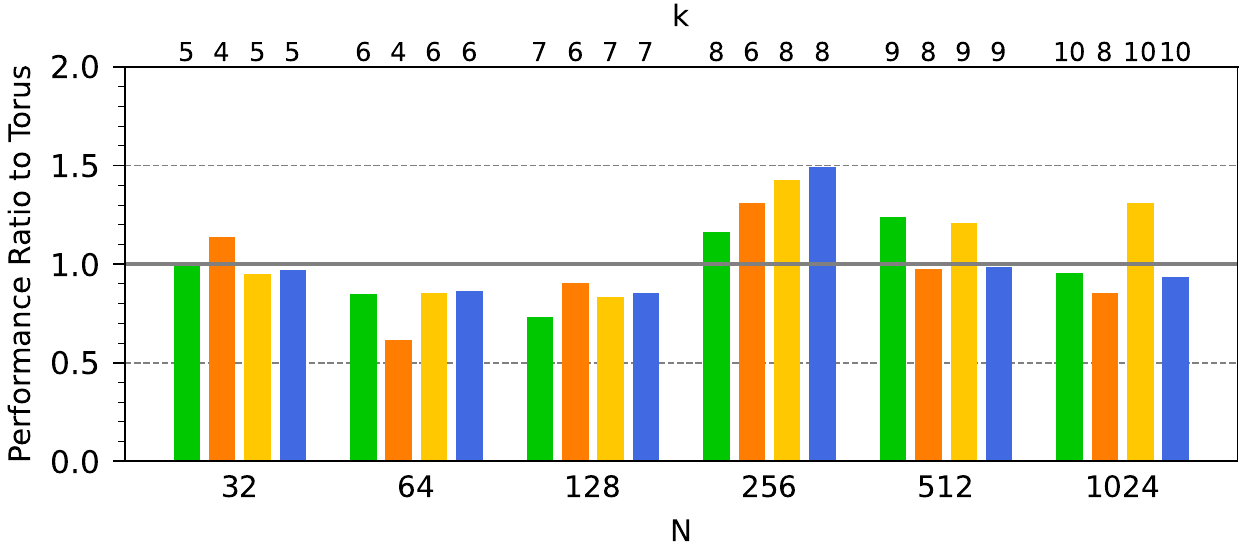}}
	\\\vspace*{0.3\baselineskip}
	\subfigure[NPB BT]{
		\includegraphics[width=0.7\textwidth]{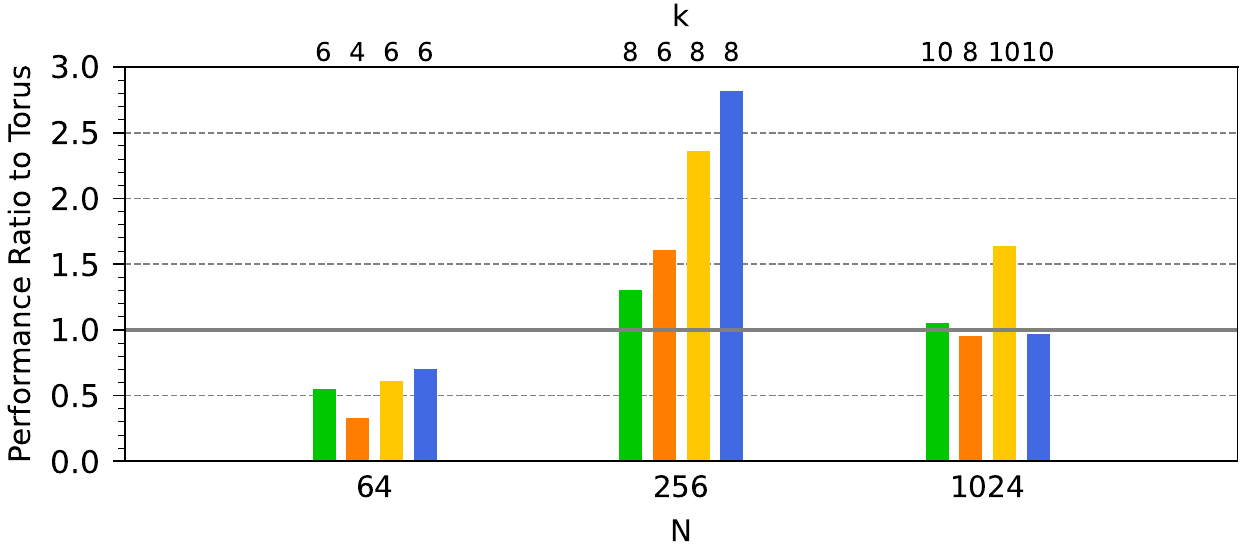}}
	\\\vspace*{0.3\baselineskip}
	\subfigure[NPB SP]{
		\includegraphics[width=0.7\textwidth]{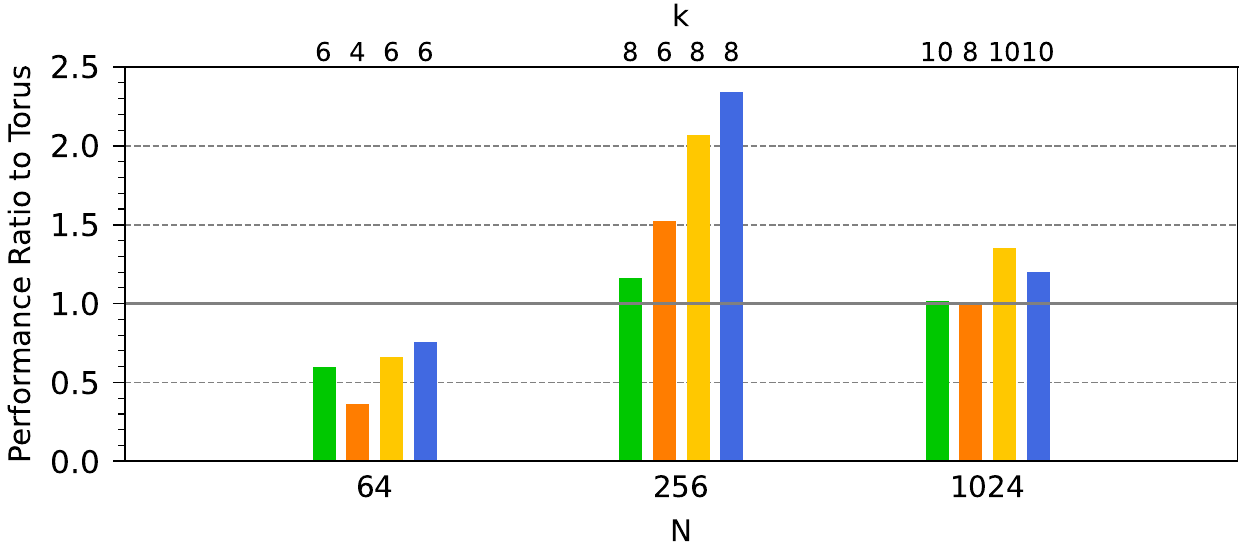}}
	\caption{Performance ratios to torus on NPB pseudo applications}
	\label{fig_npb2}
\end{figure}

The NAS Parallel Benchmarks (version 3.4.1) \cite{npb, Bailey1991} consist of programs derived from computational fluid dynamics (CFD) applications. We benchmark four kernel programs: discrete 3D FFT (FT), integer sort (IS), conjugate gradient method (CG) to calculate the smallest eigenvalue and multi-grid solver (MG) on a sequence of meshes, and three pseudo applications: lower-upper (LU) Gauss-Seidel solver, block tri-diagonal (BT) solver and scalar penta-diagonal (SP) solver. FT tests long-distance all-to-all communication; IS uses random memory access and tests both integer computation speed and communication; CG uses irregular memory access and tests unstructured long-distance communication; MG tests highly structured short- and long-distance communication with intensive memory access \cite{npb, Bailey1991}. We select the standard problem size Class C for each benchmark. Since BT and SP require a square number of processors, we perform these two applications only on topologies of $N=64$, 256 and 1024.

Figures \ref{fig_npb1} and \ref{fig_npb2} show the NPB performance ratios to torus. The performance of FT (Fig. \ref{fig_npb1}a) is similar to 1D FFTE (Fig. \ref{fig_ffte}) in accordance with global communication, in which the low-degree and high-degree optimal circulant topologies and Cartesian products all have high performance gains of 66\%, 124\% and 27\% in average respectively, with high-degree optimal circulant topologies 21\% higher than hypercube. For IS (Fig. \ref{fig_npb1}b), the performance is also similar to Graph 500 (Fig. \ref{fig_graph500}) but with relatively lower performance ratios, in which the high-degree optimal circulant topologies and Cartesian products achieve average performance gains of 30\% and 18\%. The performances of other NPB benchmarks fluctuate around the average (Fig. \ref{fig_npb1}c-d and \ref{fig_npb2}). The high-degree optimal circulant topologies are able to reach higher average performance gains of 28\%, 54\% and 36\% in CG, BT and SP, almost as good as hypercube, while maintaining around 3\% and 10\% in MG and LU. The Cartesian products achieve higher average performance gains of 35\% and 18\% in CG and MG, while keeping around equal performance in LU, BT and SP. Meanwhile, the low-degree optimal circulant topologies have almost equal average performance compared with torus in most of the NPB benchmarks apart from FT.

The NPB performance ratios to torus demonstrated multiple perspectives for performance enhancement. On one hand, applications with intensive global communication such as FT are strongly dependent on network topology, where the optimal circulant topologies contribute most to enhancing the performance. On the other hand, other factors such as data traffic patterns, internal algorithms and routing also influence the performance. One application can reach particularly high performance when its internal traffic pattern matches with the network connections, such as hypercube for IS. The Cartesian products of optimal circulant topologies and fully connected topologies may serve as better candidates for balancing the influence of global communication with specific traffic patterns, such as in CG and MG.

\section{Discussion and conclusion}\label{sec_conclude}
Optimal circulant graphs, obtained by highly efficient parallel exhaustive search algorithm in our study, are recognized as low-latency network topologies for large clusters of computing systems. The optimal circulant graphs and their Cartesian products, obtained from a huge search space of graphs with the same graph size and degree, have minimal diameter and MPL and maximum bisection width. These favorable graph properties prompt us to verify a common belief that they can significantly improve the performance for communication-intensive applications.

Indeed, the second contribution of our work is to demonstrate the enhancement of performance of these optimal circulant topologies, compared with other mainstream topologies including torus and hypercube of the same size, on effective bandwidth, FFTE, NPB FT and Graph 500. These applications not only have high communication-to-computation ratio, but also have dominant dependence on global communication, on which our newly discovered optimal circulant topologies fulfill their advantages to the most extent. We also observe that, although the optimal circulant topologies and their Cartesian products achieve higher performance than torus, the performance ratios fluctuate around the average on different topology sizes for certain benchmarks such as NPB MG, BT and SP. This exhibits the influence on performance from other factors including specific data traffic patterns, internal algorithms, topology and problem scale, routing methods and memory access. A noticeable phenomenon is the benefit of extremely symmetric hypercube for the performance of Graph 500 and NPB IS. More investigation shows that specific algorithms for MPI collective operations such as recursive doubling and halving \cite{Thakur2005} match exactly the hypercube network connections. The perfect mapping between the communication pattern and the network topology reveals another important perspective in performance enhancement which requires further exploration and may even lead to hardware-software co-design.

In addition, the Cartesian products of optimal circulant topologies and fully connected topologies have both optimal local and global structures which can balance the global communication with specific traffic patterns as shown by our simulations. As another advantage over torus and hypercube, the node degree of circulant topologies can be arbitrary. Therefore, optimal circulant topologies are ideal candidates for global connections in large-scale hierarchical networks.

\bmhead{Acknowledgments}
The authors thank Stony Brook Research Computing and Cyberinfrastructure, and the Institute for Advanced Computational Science at Stony Brook University for access to the high-performance SeaWulf computing system, which was made possible by a \$1.4M National Science Foundation grant (\#1531492).

\bibliography{references.bib}

\begin{thebibliography}{53}
\providecommand{\natexlab}[1]{#1}
\providecommand{\url}[1]{{#1}}
\providecommand{\urlprefix}{URL }
\providecommand{\doi}[1]{\url{https://doi.org/#1}}
\providecommand{\eprint}[2][]{\url{#2}}
 \bibcommenthead

\bibitem[{eff(2021)}]{effec}
 (2021) {E}ffective {B}andwidth (b\_eff) {B}enchmark.
  \url{https://fs.hlrs.de/projects/par/mpi/b\_eff/}

\bibitem[{fft(2021)}]{ffte}
 (2021) {FFTE: A Fast Fourier Transform Package}. \url{http://www.ffte.jp/}

\bibitem[{fxt(2021)}]{fxt}
 (2021) Fxt: a library of algorithms (2021). \url{https://www.jjj.de/fxt/}

\bibitem[{gra(2021)}]{graph500}
 (2021) Graph 500. \url{http://graph500.org/}

\bibitem[{npb(2021)}]{npb}
 (2021) {NPB: NAS Parallel Benchmarks}.
  \url{http://www.nas.nasa.gov/publications/npb.html}

\bibitem[{top(2021)}]{top500}
 (2021) {T}op 500 supercomputer site (2021). \url{http://www.top500.org}

\bibitem[{Ajima et~al(2009)Ajima, Sumimoto, and Shimizu}]{Ajima2009}
Ajima Y, Sumimoto S, Shimizu T (2009) Tofu: A {6D} mesh/torus interconnect for
  exascale computers. Computer 42(11):36--40. \doi{10.1109/mc.2009.370}

\bibitem[{Ajima et~al(2018)Ajima, Kawashima, Okamoto, Shida, Hirai, Shimizu,
  Hiramoto, Ikeda, Yoshikawa, Uchida, and Inoue}]{Ajima2018}
Ajima Y, Kawashima T, Okamoto T, et~al (2018) The tofu interconnect d. In: 2018
  {IEEE} International Conference on Cluster Computing ({CLUSTER}). {IEEE},
  Belfast, UK, \doi{10.1109/cluster.2018.00090}

\bibitem[{Alverson et~al(2010)Alverson, Roweth, and Kaplan}]{Alverson2010}
Alverson R, Roweth D, Kaplan L (2010) The gemini system interconnect. In: 2010
  18th {IEEE} Symposium on High Performance Interconnects. {IEEE}, Mountain
  View, CA, USA, \doi{10.1109/hoti.2010.23}

\bibitem[{Arndt(2010)}]{arndt2010matters}
Arndt J (2010) Matters Computational: ideas, algorithms, source code. Springer
  Science \& Business Media, Berlin

\bibitem[{Bailey et~al(1991)Bailey, Barszcz, Barton, Browning, Carter, Dagum,
  Fatoohi, Frederickson, Lasinski, Schreiber, Simon, Venkatakrishnan, and
  Weeratunga}]{Bailey1991}
Bailey D, Barszcz E, Barton J, et~al (1991) The {NAS} parallel benchmarks. The
  International Journal of Supercomputing Applications 5(3):63--73.
  \doi{10.1177/109434209100500306}

\bibitem[{Bermond et~al(1995)Bermond, Comellas, and Hsu}]{Bermond1995}
Bermond J, Comellas F, Hsu D (1995) Distributed loop computer-networks: A
  survey. Journal of Parallel and Distributed Computing 24(1):2--10.
  \doi{10.1006/jpdc.1995.1002},
  \urlprefix\url{https://doi.org/10.1006/jpdc.1995.1002}

\bibitem[{Bevan et~al(2017)Bevan, Erskine, and Lewis}]{Bevan2017}
Bevan D, Erskine G, Lewis R (2017) Large circulant graphs of fixed diameter and
  arbitrary degree. Ars Mathematica Contemporanea 13(2):275--291.
  \doi{10.26493/1855-3974.969.659},
  \urlprefix\url{https://doi.org/10.26493/1855-3974.969.659}

\bibitem[{Boesch and Tindell(1984)}]{Boesch1984}
Boesch F, Tindell R (1984) Circulants and their connectivities. Journal of
  Graph Theory 8(4):487--499. \doi{10.1002/jgt.3190080406},
  \urlprefix\url{https://doi.org/10.1002/jgt.3190080406}

\bibitem[{Brightwell et~al(2006)Brightwell, Pedretti, Underwood, and
  Hudson}]{Brightwell2006}
Brightwell R, Pedretti K, Underwood K, et~al (2006) {SeaStar} interconnect:
  Balanced bandwidth for scalable performance. {IEEE} Micro 26(3):41--57.
  \doi{10.1109/mm.2006.65}

\bibitem[{Casanova et~al(2014)Casanova, Giersch, Legrand, Quinson, and
  Suter}]{Casanova2014}
Casanova H, Giersch A, Legrand A, et~al (2014) Versatile, scalable, and
  accurate simulation of distributed applications and platforms. Journal of
  Parallel and Distributed Computing 74(10):2899--2917.
  \doi{10.1016/j.jpdc.2014.06.008}

\bibitem[{Cerf et~al(1974)Cerf, Cowan, Mullin, and Stanton}]{Cerf1974}
Cerf VG, Cowan DD, Mullin RC, et~al (1974) A lower bound on the average
  shortest path length in regular graphs. Networks 4(4):335--342.
  \doi{10.1002/net.3230040405}

\bibitem[{Chen et~al(2011)Chen, Parker, Eisley, Heidelberger, Senger, Sugawara,
  Kumar, Salapura, Satterfield, and Steinmacher-Burow}]{Chen2011}
Chen D, Parker JJ, Eisley NA, et~al (2011) The {IBM Blue Gene/Q}
  interconnection network and message unit. In: Proceedings of 2011
  International Conference for High Performance Computing, Networking, Storage
  and Analysis on - {SC} {\textquotesingle}11. {ACM} Press, Seattle, WA, USA,
  \doi{10.1145/2063384.2063419}

\bibitem[{Dally(1990)}]{Dally1990}
Dally W (1990) Performance analysis of k-ary n-cube interconnection networks.
  {IEEE} Transactions on Computers 39(6):775--785. \doi{10.1109/12.53599}

\bibitem[{Dally and Towles(2003)}]{Dally2004}
Dally W, Towles B (2003) Principles and Practices of Interconnection Networks.
  Elsevier Science \& Technology, Amsterdam

\bibitem[{Deng et~al(2012)Deng, Ramos, and Hornos}]{Deng2012}
Deng Y, Ramos AF, Hornos JEM (2012) Symmetry insights for design of
  supercomputer network topologies: Roots and weights lattices. International
  Journal of Modern Physics B 26(31):1250,169. \doi{10.1142/s021797921250169x}

\bibitem[{Deng et~al(2020)Deng, Guo, Ramos, Huang, Xu, and Liu}]{Deng2020}
Deng Y, Guo M, Ramos AF, et~al (2020) Optimal low-latency network topologies
  for cluster performance enhancement. The Journal of Supercomputing
  76(12):9558--9584. \doi{10.1007/s11227-020-03216-y},
  \urlprefix\url{https://doi.org/10.1007/s11227-020-03216-y}

\bibitem[{Faanes et~al(2012)Faanes, Bataineh, Roweth, Court, Froese, Alverson,
  Johnson, Kopnick, Higgins, and Reinhard}]{Faanes2012}
Faanes G, Bataineh A, Roweth D, et~al (2012) Cray cascade: A scalable {HPC}
  system based on a dragonfly network. In: 2012 International Conference for
  High Performance Computing, Networking, Storage and Analysis. {IEEE}, Salt
  Lake City, UT, USA, \doi{10.1109/sc.2012.39}

\bibitem[{Feng et~al(2012)Feng, Zhang, and Deng}]{Feng2012}
Feng R, Zhang P, Deng Y (2012) Simulated performance evaluation of a 6d
  mesh/{iBT} interconnect. In: 2012 13th {ACIS} International Conference on
  Software Engineering, Artificial Intelligence, Networking and
  Parallel/Distributed Computing. {IEEE}, Kyoto, Japan,
  \doi{10.1109/snpd.2012.19},
  \urlprefix\url{https://doi.org/10.1109/snpd.2012.19}

\bibitem[{Feng et~al(2013)Feng, Zhang, and Deng}]{Feng2013}
Feng R, Zhang P, Deng Y (2013) Deadlock-free routing algorithms for 6d
  mesh/{iBT} interconnection networks. In: 2013 14th {ACIS} International
  Conference on Software Engineering, Artificial Intelligence, Networking and
  Parallel/Distributed Computing. {IEEE}, Honolulu, HI, USA,
  \doi{10.1109/snpd.2013.43},
  \urlprefix\url{https://doi.org/10.1109/snpd.2013.43}

\bibitem[{Feria-Pur{\'{o}}n et~al(2014)Feria-Pur{\'{o}}n, Ryan, and
  P{\'{e}}rez-Ros{\'{e}}s}]{FeriaPurn2014}
Feria-Pur{\'{o}}n R, Ryan J, P{\'{e}}rez-Ros{\'{e}}s H (2014) Searching for
  large multi-loop networks. Electronic Notes in Discrete Mathematics
  46:233--240. \doi{10.1016/j.endm.2014.08.031},
  \urlprefix\url{https://doi.org/10.1016/j.endm.2014.08.031}

\bibitem[{Fu et~al(2016)Fu, Liao, Yang, Wang, Song, Huang, Yang, Xue, Liu,
  Qiao, Zhao, Yin, Hou, Zhang, Ge, Zhang, Wang, Zhou, and Yang}]{Fu2016}
Fu H, Liao J, Yang J, et~al (2016) The sunway {TaihuLight} supercomputer:
  system and applications. Science China Information Sciences 59(7).
  \doi{10.1007/s11432-016-5588-7}

\bibitem[{Gross et~al(2013)Gross, Yellen, and Zhang}]{gross2013handbook}
Gross JL, Yellen J, Zhang P (2013) Handbook of graph theory. CRC press, Boca
  Raton

\bibitem[{Hayes and Mudge(1989)}]{Hayes1989}
Hayes J, Mudge T (1989) Hypercube supercomputers. Proceedings of the {IEEE}
  77(12):1829--1841. \doi{10.1109/5.48826}

\bibitem[{Hwang(2003)}]{Hwang2003}
Hwang F (2003) A survey on multi-loop networks. Theoretical Computer Science
  299(1-3):107--121. \doi{10.1016/s0304-3975(01)00341-3},
  \urlprefix\url{https://doi.org/10.1016/s0304-3975(01)00341-3}

\bibitem[{{IBM Blue Gene Team}(2008)}]{ibm2008a}
{IBM Blue Gene Team} (2008) Overview of the {IBM Blue Gene/P} project. {IBM}
  Journal of Research and Development 52(1.2):199--220.
  \doi{10.1147/rd.521.0199}

\bibitem[{{InfiniBand$@$ Trade Association}(2016)}]{infiniband}
{InfiniBand$@$ Trade Association} (2016) {InfiniBand Architecture
  Specification, release 1.3}. URL: \url{http://wwwinfinibandtaorg}

\bibitem[{Kim et~al(2008)Kim, Dally, Scott, and Abts}]{Kim2008}
Kim J, Dally WJ, Scott S, et~al (2008) Technology-driven, highly-scalable
  dragonfly topology. In: 2008 International Symposium on Computer
  Architecture. {IEEE}, Beijing, China, \doi{10.1109/isca.2008.19}

\bibitem[{Koniges et~al(2001)Koniges, Rabenseifner, and Solchenbach}]{Konigesa}
Koniges A, Rabenseifner R, Solchenbach K (2001) Benchmark design for
  characterization of balanced high-performance architectures. In: Proceedings
  15th International Parallel and Distributed Processing Symposium. {IPDPS}
  2001. {IEEE} Comput. Soc, San Francisco, CA, USA,
  \doi{10.1109/ipdps.2001.925208}

\bibitem[{Leiserson(1985)}]{Leiserson1985}
Leiserson CE (1985) Fat-trees: Universal networks for hardware-efficient
  supercomputing. {IEEE} Transactions on Computers C-34(10):892--901.
  \doi{10.1109/tc.1985.6312192}

\bibitem[{MONAKHOVA(2012)}]{MONAKHOVA2012}
MONAKHOVA EA (2012) A survey on undirected circulant graphs. Discrete
  Mathematics, Algorithms and Applications 04(01):1250,002.
  \doi{10.1142/s1793830912500024},
  \urlprefix\url{https://doi.org/10.1142/s1793830912500024}

\bibitem[{Moore(1965)}]{Moore1965}
Moore GE (1965) Cramming more components onto integrated circuits. Electronics
  38(8):114–117

\bibitem[{Murphy et~al(2010)Murphy, Wheeler, Barrett, and
  Ang}]{murphy2010introducing}
Murphy RC, Wheeler KB, Barrett BW, et~al (2010) Introducing the graph 500. Cray
  Users Group (CUG) 19:45--74

\bibitem[{Nakao et~al(2021)Nakao, Sakai, Hanada, Murai, and Sato}]{Nakao2021}
Nakao M, Sakai M, Hanada Y, et~al (2021) Graph optimization algorithm for
  low-latency interconnection networks. Parallel Computing 106:102,805.
  \doi{10.1016/j.parco.2021.102805},
  \urlprefix\url{https://doi.org/10.1016/j.parco.2021.102805}

\bibitem[{Ruskey(2003)}]{ruskey2003combinatorial}
Ruskey F (2003) Combinatorial generation. Preliminary working draft University
  of Victoria, Victoria, BC, Canada 11:20

\bibitem[{Sabino et~al(2018)Sabino, Vasconcelos, Deng, and Ramos}]{Sabino2018}
Sabino AU, Vasconcelos MFS, Deng Y, et~al (2018) Symmetry-guided design of
  topologies for supercomputer networks. International Journal of Modern
  Physics C 29(07):1850,048. \doi{10.1142/s0129183118500481}

\bibitem[{Sanders and Schulz(2013)}]{Sanders2013}
Sanders P, Schulz C (2013) Think locally, act globally: Highly balanced graph
  partitioning. In: Experimental Algorithms. Lecture notes in computer science,
  Springer Berlin Heidelberg, Berlin, Heidelberg, p 164--175,
  \doi{10.1007/978-3-642-38527-8\_16}

\bibitem[{Sensi et~al(2020)Sensi, Girolamo, McMahon, Roweth, and
  Hoefler}]{DeSensi2020}
Sensi DD, Girolamo SD, McMahon KH, et~al (2020) An in-depth analysis of the
  slingshot interconnect. In: {SC}20: International Conference for High
  Performance Computing, Networking, Storage and Analysis. {IEEE}, Atlanta, GA,
  USA, \doi{10.1109/sc41405.2020.00039},
  \urlprefix\url{https://doi.org/10.1109/sc41405.2020.00039}

\bibitem[{Takahashi and Kanada(2000)}]{Takahashi2000}
Takahashi D, Kanada Y (2000) High-performance radix-2, 3 and 5 parallel {1-D}
  complex {FFT} algorithms for distributed-memory parallel computers. The
  Journal of Supercomputing 15(2):207--228. \doi{10.1023/a:1008160021085}

\bibitem[{Thakur et~al(2005)Thakur, Rabenseifner, and Gropp}]{Thakur2005}
Thakur R, Rabenseifner R, Gropp W (2005) Optimization of collective
  communication operations in {MPICH}. The International Journal of High
  Performance Computing Applications 19(1):49--66.
  \doi{10.1177/1094342005051521},
  \urlprefix\url{https://doi.org/10.1177/1094342005051521}

\bibitem[{Truong et~al(2017)Truong, Fujiwara, Koibuchi, and
  Nguyen}]{Truong2017}
Truong NT, Fujiwara I, Koibuchi M, et~al (2017) Distributed shortcut networks:
  Low-latency low-degree non-random topologies targeting the diameter and cable
  length trade-off. {IEEE} Transactions on Parallel and Distributed Systems
  28(4):989--1001. \doi{10.1109/tpds.2016.2613043},
  \urlprefix\url{https://doi.org/10.1109/tpds.2016.2613043}

\bibitem[{Xu et~al(2019)Xu, Huang, Jimenez, and Deng}]{Xu2019}
Xu Z, Huang X, Jimenez F, et~al (2019) A new record of graph enumeration
  enabled by parallel processing. Mathematics 7(12):1214.
  \doi{10.3390/math7121214}

\bibitem[{Yasudo et~al(2019)Yasudo, Koibuchi, Nakano, Matsutani, and
  Amano}]{Yasudo2019}
Yasudo R, Koibuchi M, Nakano K, et~al (2019) Designing high-performance
  interconnection networks with host-switch graphs. {IEEE} Transactions on
  Parallel and Distributed Systems 30(2):315--330.
  \doi{10.1109/tpds.2018.2864286},
  \urlprefix\url{https://doi.org/10.1109/tpds.2018.2864286}

\bibitem[{Zhang and Deng(2012{\natexlab{a}})}]{Zhang2012_ibt}
Zhang P, Deng Y (2012{\natexlab{a}}) An analysis of the topological properties
  of the interlaced bypass torus ({iBT}) networks. Applied Mathematics Letters
  25(12):2147--2155. \doi{10.1016/j.aml.2012.05.013},
  \urlprefix\url{https://doi.org/10.1016/j.aml.2012.05.013}

\bibitem[{Zhang and Deng(2012{\natexlab{b}})}]{Zhang2012_ibt_bcast}
Zhang P, Deng Y (2012{\natexlab{b}}) Design and analysis of pipelined broadcast
  algorithms for the all-port interlaced bypass torus networks. {IEEE}
  Transactions on Parallel and Distributed Systems 23(12):2245--2253.
  \doi{10.1109/tpds.2012.93},
  \urlprefix\url{https://doi.org/10.1109/tpds.2012.93}

\bibitem[{Zhang et~al(2011)Zhang, Powell, and Deng}]{Zhang2011}
Zhang P, Powell R, Deng Y (2011) Interlacing bypass rings to torus networks for
  more efficient networks. {IEEE} Transactions on Parallel and Distributed
  Systems 22(2):287--295. \doi{10.1109/tpds.2010.89}

\bibitem[{Zhang et~al(2015)Zhang, Deng, Feng, Luo, and Wu}]{Zhang2015}
Zhang P, Deng Y, Feng R, et~al (2015) Evaluation of various networks
  configurated by adding bypass or torus links. {IEEE} Transactions on Parallel
  and Distributed Systems 26(4):984--996. \doi{10.1109/tpds.2014.2315201}

\bibitem[{Zhang et~al(2019)Zhang, Huang, Xu, and Deng}]{Zhang2019}
Zhang Y, Huang X, Xu Z, et~al (2019) A structured table of graphs with
  symmetries and other special properties. Symmetry 12(1):2.
  \doi{10.3390/sym12010002},
  \urlprefix\url{https://doi.org/10.3390/sym12010002}

\end{thebibliography}

\end{document}